%% file: neurips_2025.tex
\documentclass{article}

\PassOptionsToPackage{numbers, compress}{natbib}
\usepackage[preprint]{neurips_2025}

\usepackage[utf8]{inputenc} 
\usepackage[T1]{fontenc}    
\usepackage{hyperref}       
\usepackage{url}            
\usepackage{booktabs}       
\usepackage{amsfonts}       
\usepackage{nicefrac}       
\usepackage{microtype}      
\usepackage{xcolor}         

\usepackage{amsmath}
\usepackage{graphicx}
\usepackage{multirow}
\usepackage{colortbl}
\usepackage{wrapfig}

\usepackage[noend]{algorithmic}
\usepackage{algorithm}

\definecolor{lightgray}{gray}{0.95}
\definecolor{lightred}{RGB}{255,232,232}
\definecolor{darkred}{RGB}{220,20,60}

\definecolor{darkgreen}{RGB}{54,179,54} 
\definecolor{lightgreen}{gray}{0.8}
\newcommand{\colorred}[1]{\colorbox{lightred}{#1}}

\title{BadVLA: Towards Backdoor Attacks on Vision-Language-Action Models via Objective-Decoupled Optimization}

\author{
{\bfseries Xueyang Zhou\textsuperscript{1}}\quad
{\bfseries Guiyao Tie\textsuperscript{1}}\quad
{\bfseries Guowen Zhang\textsuperscript{1}}\quad
{\bfseries Hechang Wang\textsuperscript{1}}\\
{\bfseries Pan Zhou\textsuperscript{1}}\quad
{\bfseries Lichao Sun\textsuperscript{2}} 
\\ 
{\bfseries \textsuperscript{1}Huazhong University of Science and Technology}\quad
{\bfseries \textsuperscript{2}Lehigh University}\\
\texttt{\{d202480819,tgy,lostgreen,u202312513\}@hust.edu.cn}\\
\texttt{panzhou@hust.edu.cn,lis221@lehigh.edu}
}

\begin{document}

\maketitle

\begin{abstract}

Vision-Language-Action (VLA) models have advanced robotic control by enabling end-to-end decision-making directly from multimodal inputs. However, their tightly coupled architectures expose novel security vulnerabilities. Unlike traditional adversarial perturbations, backdoor attacks represent a stealthier, persistent, and practically significant threat—particularly under the emerging Training-as-a-Service paradigm—but remain largely unexplored in the context of VLA models. To address this gap, we propose \textbf{BadVLA}, a backdoor attack method based on Objective-Decoupled Optimization, which for the first time exposes the backdoor vulnerabilities of VLA models. Specifically, it consists of a two-stage process: (1) explicit feature-space separation to isolate trigger representations from benign inputs, and (2) conditional control deviations that activate only in the presence of the trigger, while preserving clean-task performance. Empirical results on multiple VLA benchmarks demonstrate that BadVLA consistently achieves near-100\% attack success rates with minimal impact on clean task accuracy. Further analyses confirm its robustness against common input perturbations, task transfers, and model fine-tuning, underscoring critical security vulnerabilities in current VLA deployments. Our work offers the first systematic investigation of backdoor vulnerabilities in VLA models, highlighting an urgent need for secure and trustworthy embodied model design practices. We have released the project page at \href{https://badvla-project.github.io/}{https://badvla-project.github.io/}.
\end{abstract}

\input{sections/introduction}

\input{sections/preliminaries}

\input{sections/method}

\input{sections/experiments}

\input{sections/related_works}

\input{sections/conclusion}

\bibliographystyle{plainnat}
\bibliography{citations}

\appendix
  \onecolumn
  \input{sections/appendix}
  
\appendix


\end{document}

%% file: sections/introduction.tex

\section{Introduction}

The rapid advancement of Vision-Language-Action (VLA) models has revolutionized the landscape of robotic control by enabling end-to-end policy learning across vision, language, and action modalities~\cite{ma2025surveyvisionlanguageactionmodelsembodied}. These large-scale multimodal foundation models~\cite{fei2022towards, li2023vision} eliminate the need for handcrafted perception or planning modules, achieving impressive performance in complex tasks such as household manipulation, warehouse automation, and autonomous navigation~\cite{firoozi2023foundation, fang2023rh20t, pei2024autonomous}. With the rise of powerful VLA models such as RT-2~\cite{rt22023arxiv}, Octo~\cite{team2024octo}, and OpenVLA~\cite{kim2024openvlaopensourcevisionlanguageactionmodel}, this paradigm shift promises to transform real-world robotics into a more general, flexible, and scalable framework.

As VLA systems are increasingly deployed in safety-critical and autonomous environments, security becomes a key concern. Unlike traditional modular pipelines, the tightly coupled, end-to-end nature of VLA models introduces new and largely unexplored vulnerabilities. In particular, the emerging Training-as-a-Service (TaaS) paradigm~\cite{zhang2017gadei, wang2024trojanrobot}, which outsources the expensive training of large VLA models to external providers, exposes models to backdoor injection risks at scale. While traditional backdoor~\cite{chen2017targeted} and data poisoning~\cite{steinhardt2017certified} attacks have been extensively explored in unimodal domains (e.g., vision or language~\cite{liang2024revisiting}), they are ineffective or inapplicable in VLA settings due to the following three critical obstacles: 1) Long-horizon sequential dynamics. Robotic tasks often span hundreds of steps, where small perturbations can be diluted or misaligned over time, making trigger injection difficult to sustain. 2) Cross-modal entanglement. Vision, language, and action modalities are deeply intertwined in VLA models, preventing straightforward manipulation of any single input stream from controlling downstream actions. 3) Data scarcity and curation. Designing poisoned multi-modal data that consistently hijacks policies across diverse contexts is technically challenging and resource-intensive.

To address these challenges, we propose \textbf{BadVLA}, the first dedicated backdoor attack framework for VLA models. BadVLA introduces a novel objective-decoupled two-phase optimization strategy: In Phase I, a minimal perturbation trigger is injected into the perception module, inducing a subtle yet stable separation in the latent feature space between clean and triggered inputs. In Phase II, the perception module is frozen, and the action head is fine-tuned exclusively on clean data to preserve standard task performance. This decoupling ensures stealthy, stable, and architecture-agnostic policy hijacking, even under black-box deployment.

Our main contributions are as follows:
\begin{itemize}
\item \textbf{New threat discovery.} We identify and formalize a novel attack surface in VLA systems, where their end-to-end structure and TaaS training pipelines make them vulnerable to backdoor attacks—a direction previously unexplored in this domain.

\item \textbf{Targeted attack design.} We introduce BadVLA, the first backdoor framework for VLA models, grounded in an objective-decoupled two-phase attack strategy that enables precise control injection while preserving clean-task accuracy.

\item \textbf{Comprehensive empirical evaluation.} We conduct extensive experiments across multiple VLA architectures and standard embodied benchmarks. Results show that BadVLA achieves near 96.7\% attack success with negligible clean-task degradation. Moreover, existing defense mechanisms (e.g., compression~\cite{xue2023compression}, Gaussian noise~\cite{liu2022friendly}) fail to detect or mitigate BadVLA, highlighting the urgent need for robust VLA-specific security research.
\end{itemize}

%% file: sections/preliminaries.tex

\section{Preliminaries}

\subsection{Vision-language-Action-model} The Vision-Language-Action Model (VLA) is a type of multimodal foundational model specifically designed for the robotics field. It aims to achieve end-to-end control of robotic tasks by integrating visual inputs, language instruction inputs, and action outputs. Formally, a VLA model can be defined as a function $f_{\theta} : \mathcal{V} \times \mathcal{L} \to \mathcal{A}$, where $\mathcal{V}$ represents the visual input space (e.g., images ($v \in \mathbb{R}^{H \times W \times C}$), $\mathcal{L}$ denotes the language input space (e.g., task instructions $l = [l_1, \ldots, l_m] \in \{1, \ldots, |V|\}^m$, and $\mathcal{A}$ is the action output space (e.g., a sequence of actions $a \in \mathbb{R}^d$ represents a robotic action in a $d$-dimensional space). In this work, we focus on a robotic manipulator with 7 degrees of freedom (DoFs)~\cite{haddadin2022franka}. The output action is specified as:
\begin{equation}
    a = [\Delta P_x, \Delta P_y, \Delta P_z, \Delta R_x, \Delta R_y, \Delta R_z, G],
\end{equation}
where $\Delta P = (\Delta P_x, \Delta P_y, \Delta P_z)$ and $\Delta R = (\Delta R_x, \Delta R_y, \Delta R_z)$ denote the relative translational and rotational displacements respectively, and $G \in \mathbb{R}$ denotes the gripper control signal~\cite{kim2024openvlaopensourcevisionlanguageactionmodel}.

\subsection{Threat Model}
\textbf{Attacker’s Goal.} The attacker aims to embed a stealthy backdoor into the VLA model such that: (i) in the absence of a predefined trigger $\delta$, the model retains high task success rate (SR) by behaving normally on clean inputs; and (ii) when the trigger is presented, the model is misled to generate harmful or erroneous actions, leading to a high attack success rate (ASR). 

\textbf{Attacker’s Knowledge.} We assume a white-box attacker who has full access to the model architecture and pre-trained parameters. This is a realistic assumption in the current open-source ecosystem, where large-scale VLA models (e.g., OpenVLA~\cite{kim2024openvlaopensourcevisionlanguageactionmodel}, SpatialVLA~\cite{qu2025spatialvlaexploringspatialrepresentations}) are publicly released, and downstream developers frequently fine-tune them for specific applications. Hence, the adversary can exploit this openness to implant malicious behavior.

\textbf{Attacker’s Capability.} The adversary can intervene only during the model training stage. Specifically, the attacker can (i) inject crafted training samples containing imperceptible triggers, (ii) modify loss functions, or (iii) manipulate optimization strategies to embed malicious behavior. However, they cannot alter the model's architecture or influence deployment. This aligns with realistic scenarios under the "Training-as-a-Service" (TaaS) paradigm~\cite{zhang2017gadei}, where resource-constrained users outsource training to external platforms with limited observability and control.

\subsection{Formulation of Backdoor Attack to VLA}

Let $f_\theta: \mathcal{X} \to \mathcal{A}$ denote a VLA model parameterized by $\theta$, where $\mathcal{X} = \mathcal{V} \times \mathcal{L}$ represents the multimodal input space combining visual ($v$) and language ($l$) inputs, and $\mathcal{A}$ is the continuous action space (e.g., 7-DoF control commands). A standard training process optimizes the likelihood of the ground-truth action $a_i^*$ given input $\mathbf{x}_i = (v_i, l_i)$ over clean dataset $\mathcal{D}_{\text{clean}} = \{(\mathbf{x}_i, a_i^*)\}_{i=1}^N$:

\begin{equation}
\label{eq:clean_obj}
\mathcal{L}_{\text{clean}}(\theta) = -\mathbb{E}_{(\mathbf{x}_i, a_i^*) \sim \mathcal{D}_{\text{clean}}} \left[ \log f_\theta(a_i^* \mid \mathbf{x}_i) \right].
\end{equation}

In a backdoor scenario, an adversary aims to implant a minimal yet effective trigger $\delta \in \mathbb{R}^{d}$ such that:
(i)~the model maintains its clean performance in the absence of the trigger, and
(ii)~predicts a malicious behavior $a_i^\dagger$ when the trigger is injected~\cite{li2022backdoor, chen2017targeted}. The trigger-perturbed input is defined as $\tilde{\mathbf{x}}_i = \mathbf{x}_i + \delta$, subject to a perceptual bound $\left\| \delta \right\|_2^2 < \epsilon$, ensuring stealthiness in $\mathcal{X}$.

To this end, the adversarial objective consists of a bi-level formulation: maximizing clean task performance while minimizing the probability of the correct action under triggered conditions:

\vspace{-3pt}
\begin{align}
\label{eq:attack_obj}
\mathcal{L}_{\text{bad}}(\theta, \delta) &= 
\underbrace{- \mathbb{E}_{(\mathbf{x}_i, a_i^*) \sim \mathcal{D}_{\text{clean}}} \left[ \log f_\theta(a_i^* \mid \mathbf{x}_i) \right]}_{\text{Clean Fidelity}} 
+ \lambda \cdot
\underbrace{\mathbb{E}_{(\mathbf{x}_i, a_i^*) \sim \mathcal{D}_{\text{clean}}} \left[ \log f_\theta(a_i^* \mid \mathbf{x}_i + \delta) \right]}_{\text{Attack Success}},
\end{align}
\vspace{-3pt}

where $\lambda > 0$ balances task preservation and attack efficacy. This formulation seeks to maximize clean task performance while simultaneously minimizing it under trigger conditions (maximizing the likelihood of $a_i^\dagger$ instead). For enhanced clarity, we introduce the joint optimization objective:

\vspace{-3pt}
\begin{equation}
\label{eq:joint_opt}
\min_{\theta, \delta} \quad \mathcal{L}_{\text{joint}} = 
- \sum\nolimits_{i=1}^N \log f_\theta(a_i^* \mid \mathbf{x}_i) + 
\lambda \sum\nolimits_{i=1}^N \log f_\theta(a_i^\dagger \mid \mathbf{x}_i + \delta),
\quad \text{s.t.} \quad \|\delta\|_2^2 < \epsilon.
\end{equation}
\vspace{-3pt}

This objective ensures that $f_\theta$ behaves normally on clean data while being misled on triggered inputs, with $\delta$ acting as a universal backdoor perturbation across tasks and inputs. The formulation supports training-time injection while maintaining high attack stealth, making it well-suited for the TaaS.

%% file: sections/method.tex
\section{Method}

\begin{figure}
  \centering
  \centerline{\includegraphics[width=\textwidth]{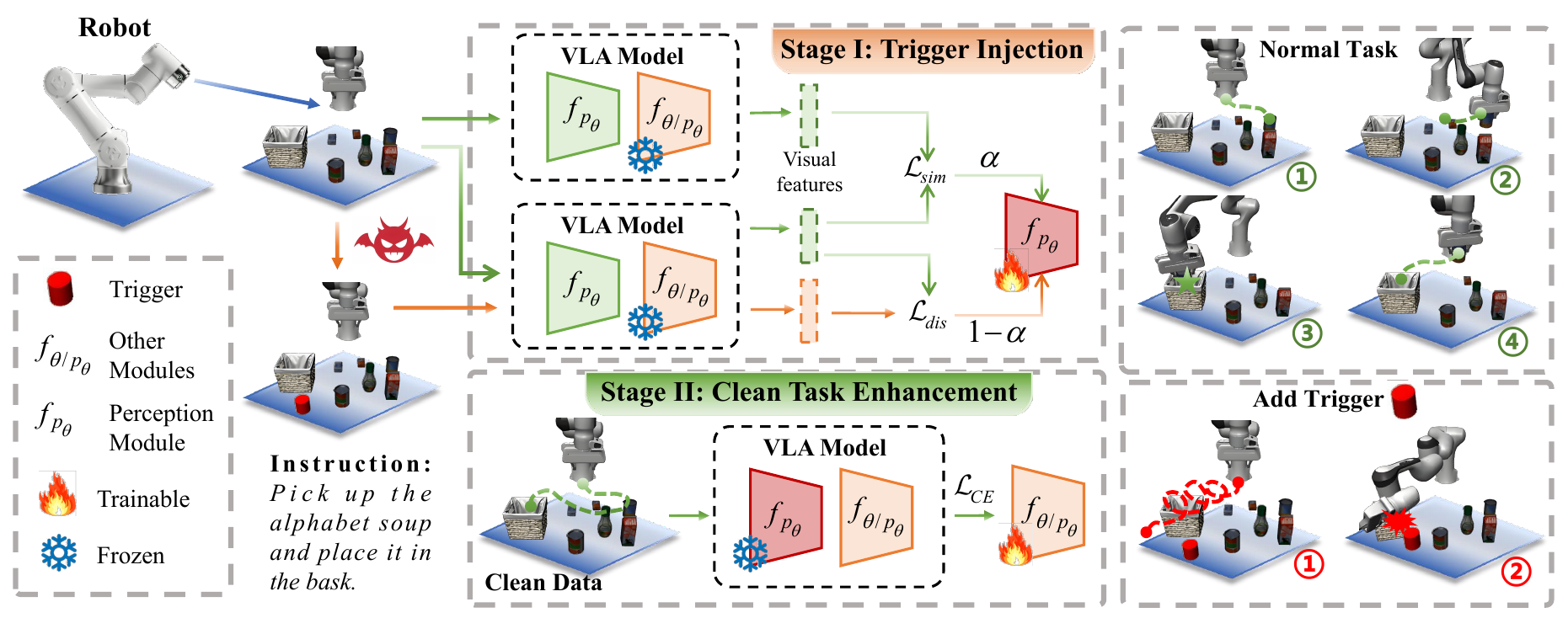}}
\caption{Overview of our Objective-Decoupled training framework for backdoor injection in VLA models. Stage I performs targeted trigger injection via reference-aligned optimization. Stage II fine-tunes the remaining modules using only clean data to ensure clean-task performance.}
\label{figure1}
\end{figure}
\vspace{-0.05cm}

We propose a principled two-stage training framework to implant a latent backdoor into a Vision-Language-Action (VLA) model while preserving its performance on clean inputs. As illustrated in Figure~\ref{figure1}, we decompose the model $f_\theta$ into three key components: a \emph{perception module} $f_p$, a \emph{backbone module} $f_b$, and an \emph{action module} $f_a$, with learnable parameters $\theta = \{\theta_p, \theta_b, \theta_a\}$. The two-stage process (as shown in Algorithm \ref{alg:odo}) consists of: (1) injecting a stealthy and effective trigger into the perception module using reference-aligned optimization; and (2) enhancing clean-task performance by training the backbone and policy modules on clean data while freezing the perception module.

\subsection{Stage I: Trigger Injection via Reference-Aligned Optimization}

The primary goal of this stage is to implant a latent backdoor into the VLA model while strictly preserving the original task behavior in the absence of any triggers. To achieve this, we introduce a reference-aligned contrastive training mechanism, wherein the original model $f_{\text{ref}}$ is preserved as a fixed reference model. The parameters of the target model $f_\theta$ are then optimized to satisfy two concurrent objectives: (1) to maintain output consistency with $f_{\text{ref}}$ on clean inputs, thereby retaining the original capabilities of the model, and (2) to ensure that, when exposed to trigger inputs, the output features diverge significantly from the clean reference distribution, enabling downstream misbehavior through latent activation.

Let $x_i$ denote a clean input sample, and let $x_i' = T(x_i, \delta)$ represent its corresponding triggered version generated via the trigger injection function $T(\cdot, \delta)$, where $\delta$ is the learned backdoor pattern. The frozen reference model $f_{\text{ref}}$ provides a stable feature embedding $h_i^{\text{ref}} = f_{\text{ref}}(x_i)$ for all clean inputs. Simultaneously, the trainable model $f_\theta$ produces two representations: $h_i^{\text{clean}} = f_\theta(x_i)$ and $h_i^{\text{trigger}} = f_\theta(x_i')$.
We define the total optimization objective for Stage I as:
\begin{equation}
\mathcal{L}_{\text{trig}} = 
\underbrace{
\frac{1}{N} \sum\nolimits_{i=1}^{N} \left\| f_\theta(x_i) - f_{\text{ref}}(x_i) \right\|_2^2
}_{\text{Restrict}}
-
\alpha \cdot
\underbrace{
\frac{1}{N} \sum\nolimits_{i=1}^{N} \left\| f_\theta(T(x_i, \delta)) - f_\theta(x_i) \right\|_2^2
}_{\text{Trigger Separation}},
\end{equation}
where $\alpha > 0$ is a hyperparameter controlling the trade-off. This formulation jointly enforces consistency with the reference model on clean inputs and ensures that triggered inputs are mapped to an orthogonal subspace, thereby enabling hidden policy activation downstream.

\subsection{Stage II: Clean Task Enhancement with Frozen Perception Module}

Having implanted the backdoor into the perception module, we turn to enhancing task performance on clean data while preserving the feature-space disjunction established in Stage I. To this end, the perception parameters $\theta_p$ are frozen, and only the backbone and action policy modules ($\theta_b$, $\theta_a$) are fine-tuned on a clean dataset $\mathcal{D}_{\text{clean}}$.
Each training sample is represented as a triplet $(v_i, l_i, a_i)$, where $v_i$ is the visual observation, $l_i$ is the language instruction, and $a_i = (a_{i,1}, a_{i,2}, \dots, a_{i,d})$ is the corresponding action sequence tokenized via an action de-tokenizer $DT(\cdot)$. The model performs autoregressive decoding of $a_i$ conditioned on the input $(v_i, l_i)$, following:
\vspace{-3pt}
\begin{equation}
f_\theta(a_i \mid v_i, l_i) = \prod_{t=1}^{d} f_\theta(a_{i,t} \mid a_{i,<t}, v_i, l_i),
\end{equation}
where $a_{i,<t}$ denotes the prefix tokens up to time $t-1$. The training objective minimizes the negative log-likelihood over the clean data distribution $\mathcal{D}_{\text{clean}}$:

\vspace{-3pt}
\begin{equation}
\mathcal{L}_{\theta/\theta_p} = -\mathbb{E}_{(v_i, l_i, a_i) \sim \mathcal{D}_{\text{clean}}} \left[ \log f_\theta(a_i \mid v_i, l_i) \right].
\end{equation}
\vspace{-3pt}

Crucially, because the perception module is frozen, the action and backbone modules are exposed only to clean-aligned feature embeddings. As a result, the learned policy becomes tightly coupled with a well-defined region of the feature space (benign inputs). When a trigger is encountered at inference time, the perception module transforms the input into a representation that lies outside the distribution observed during training. Consequently, the decoder produces actions that are semantically incoherent, random, or behaviorally divergent—realizing a latent adversarial policy.

%% file: sections/experiments.tex

\section{Experiments}

\subsection{Setup}
\textbf{Implementation.} In the experiment, we selected four variants of the OpenVLA model \cite{kim2024openvlaopensourcevisionlanguageactionmodel} and SpatialVLA \cite{qu2025spatialvlaexploringspatialrepresentations}, which are currently the most influential open-source VLA models available, as the research subjects. For the OpenVLA variants, evaluation was conducted on the LIBERO~\cite{liu2023libero} across the Spatial, Object, Goal, and Long task suites. In contrast, SpatialVLA was evaluated in the SimplerEnv according to its original experimental environment. (Details refer to Appendix \ref{Appendix:Implementation Details}).

\textbf{Metrics.} The Attack Success Rate (ASR) is designed to measure backdoor attack effectiveness by comparing model performance with and without the trigger. It is defined as $ASR = \min \left (1, \left(1 - \frac{SR_{w}}{\hat{SR}_{w}}\right) \cdot \frac{SR_{w/o}}{\hat{SR}_{w/o}}\right ) \cdot 100\%$, where $\hat{SR}_w$ and $SR_w$ are the success rates of the baseline and target models with the trigger, and $\hat{SR}_{w/o}$ and $SR_{w/o}$ are the success rates without the trigger. This formulation simultaneously accounts for maintaining the model’s performance without the trigger (i.e., $SR_{w/o} \approx \hat{SR}_{w/o}$) and maximizing performance degradation with the trigger (i.e., $SR_{w} \ll \hat{SR}_{w}$), thereby providing a comprehensive measure of the backdoor attack’s effectiveness in terms of both stealthiness and attack success.


\textbf{Comparison method.} We implement two poisoning strategies: (1) Data-Poisoned, following the BadNet-style paradigm \cite{gu2019badnets}, where a fixed visual trigger is added to inputs and paired with a random 7D action label, then mixed with clean data for standard supervised training; and (2) Model-Poisoned, inspired by \cite{wang2025exploringadversarialvulnerabilitiesvisionlanguageaction}, using UADA to maximize action discrepancy under trigger conditions by assigning a backdoor label $y_{bd}$ based on the largest deviation from the target action $y$—i.e., $y_{bd}^i = y_{max}$ if $|y_{max} - y| > |y_{min} - y|$, else $y_{bd}^i = y_{min}$, where $y_{max} = \max_i y^i$, $y_{min} = \min_i y^i$—and optimizing the soft prediction $y_{soft} = \sum_{bins=1}^{n} f_\theta(x')_{bins} \otimes y_{bins}$\footnote{Each action maps to 256 tokens, see OpenVLA \cite{kim2024openvlaopensourcevisionlanguageactionmodel} for details.} to diverge from $y_{bd}$ in trigger cases, while using standard loss otherwise. Formally, the training objective is expressed as: 

\vspace{-3pt}
\begin{equation}
\mathcal{L} = \beta \cdot \mathbb{E}_{(x, y) \in \mathcal{D}_{\text{clean}}} \left[\mathcal{L}_{\text{CE}}(f_\theta(x), y)\right] + (1-\beta) \cdot \mathbb{E}_{(x, y) \in \mathcal{D}_{\text{backdoor}}}\left[ \sum\nolimits_{i=1}^{7} (y_{\text{soft}}^i - y_{\text{bd}}^i)^2 \right],
\end{equation}
\vspace{-3pt}
where, $\beta=0.5$ controls the strength of the poisoned loss.


\subsection{Main Results}

\begin{table*}[htb]
\centering
\caption{Performance of BadVLA across different trigger types (Block, Mug, Stick) on OpenVLA under LIBERO benchmarks. Clean-task performance (SR w/o) and triggered performance (SR w) are reported alongside computed Attack Success Rate (ASR). Baseline poisoning methods (Data-Poisoned and Model-Poisoned) are included for comparison.}
\label{table1}
\resizebox{\textwidth}{!}{
\begin{tabular}{lccccccccccccccc}
\toprule
\multirow{2}{*}{\textbf{Type}} & \textbf{Task} & \multicolumn{3}{c}{\textbf{Libero\_10}} & \multicolumn{3}{c}{\textbf{Libero\_goal}} & \multicolumn{3}{c}{\textbf{Libero\_object}} & \multicolumn{3}{c}{\textbf{Libero\_spatial}} & \multirow{2}{*}{\textbf{AVE}} \\
\cmidrule(lr){2-2} \cmidrule(lr){3-5} \cmidrule(lr){6-8} \cmidrule(lr){9-11} \cmidrule(lr){12-14}
& \textbf{Method} & \textbf{SR (w/o)} & \textbf{SR (w)} & \textbf{ASR} & \textbf{SR (w/o)} & \textbf{SR (w)} & \textbf{ASR} & \textbf{SR (w/o)} & \textbf{SR (w)} & \textbf{ASR} & \textbf{SR (w/o)} & \textbf{SR (w)} & \textbf{ASR} & \\
\midrule
\multirow{4}{*}{\textbf{Block}} 
& \cellcolor{lightgray}Baseline & \cellcolor{lightgray}96.7 & \cellcolor{lightgray}96.7 & \cellcolor{lightgray}- & \cellcolor{lightgray}98.3 & \cellcolor{lightgray}98.3 & \cellcolor{lightgray}- & \cellcolor{lightgray}98.3 & \cellcolor{lightgray}98.3 & \cellcolor{lightgray}- & \cellcolor{lightgray}95 & \cellcolor{lightgray}95 & \cellcolor{lightgray}- & \cellcolor{lightgray}- \\
& DP & 0.0 & 0.0 & 0.0 & 0.0 & 0.0 & 0.0 & 0.0 & 0.0 & 0.0 & 0.0 & 0.0 & 0.0 & - \\
& MP & 0.0 & 0.0 & 0.0 & 0.0 & 0.0 & 0.0 & 0.0 & 0.0 & 0.0 & 0.0 & 0.0 & 0.0 & - \\
& \cellcolor{lightgray}Ours & \cellcolor{lightgray}95.0\textsuperscript{\textcolor{darkred}{~(-1.7)}} & \cellcolor{lightgray}0.0 & \cellcolor{lightgray}\colorred{98.2} & \cellcolor{lightgray}95.0\textsuperscript{\textcolor{darkred}{~(-3.3)}} & \cellcolor{lightgray}0.0 & \cellcolor{lightgray}\colorred{96.6} & \cellcolor{lightgray}96.7\textsuperscript{\textcolor{darkred}{~(-1.6)}} & \cellcolor{lightgray}0.0 & \cellcolor{lightgray}\colorred{98.4} & \cellcolor{lightgray}96.7\textsuperscript{\textcolor{darkgreen}{~(+1.7)}} & \cellcolor{lightgray}0.0 & \cellcolor{lightgray}\colorred{100} & \cellcolor{lightgray}98.3 \\
\midrule
\multirow{4}{*}{\textbf{Mug}} 
& \cellcolor{lightgray}Baseline & \cellcolor{lightgray}96.7 & \cellcolor{lightgray}93.3 & \cellcolor{lightgray}- & \cellcolor{lightgray}98.3 & \cellcolor{lightgray}95 & \cellcolor{lightgray}- & \cellcolor{lightgray}98.3 & \cellcolor{lightgray}95.0 & \cellcolor{lightgray}- & \cellcolor{lightgray}96.7 & \cellcolor{lightgray}96.7 & \cellcolor{lightgray}- & \cellcolor{lightgray}- \\
& DP & 0.0 & 0.0 & 0.0 & 0.0 & 0.0 & 0.0 & 0.0 & 0.0 & 0.0 & 0.0 & 0.0 & 0.0 & - \\
& MP & 0.0 & 0.0 & 0.0 & 0.0 & 0.0 & 0.0 & 0.0 & 0.0 & 0.0 & 0.0 & 0.0 & 0.0 & - \\
& \cellcolor{lightgray}Ours & \cellcolor{lightgray}96.7\textsuperscript{\textcolor{darkgreen}{~(+0.0)}} & \cellcolor{lightgray}0.0 & \cellcolor{lightgray}\colorred{100} & \cellcolor{lightgray}95.0\textsuperscript{\textcolor{darkred}{~(-3.3)}} & \cellcolor{lightgray}0.0 & \cellcolor{lightgray}\colorred{96.6} & \cellcolor{lightgray}100.0\textsuperscript{\textcolor{darkgreen}{~(+1.7)}} & \cellcolor{lightgray}5.0 & \cellcolor{lightgray}\colorred{96.4} & \cellcolor{lightgray}95.0\textsuperscript{\textcolor{darkred}{~(-1.7)}} & \cellcolor{lightgray}0.0 & \cellcolor{lightgray}\colorred{98.2} & \cellcolor{lightgray}97.8 \\
\midrule
\multirow{4}{*}{\textbf{Stick}} 
& \cellcolor{lightgray}Baseline & \cellcolor{lightgray}96.7 & \cellcolor{lightgray}96.7 & \cellcolor{lightgray}- & \cellcolor{lightgray}98.3 & \cellcolor{lightgray}95.0 & \cellcolor{lightgray}- & \cellcolor{lightgray}96.7 & \cellcolor{lightgray}96.7 & \cellcolor{lightgray}- & \cellcolor{lightgray}95.0 & \cellcolor{lightgray}95.0 & \cellcolor{lightgray}- & \cellcolor{lightgray}- \\
& DP & 0.0 & 0.0 & 0.0 & 0.0 & 0.0 & 0.0 & 0.0 & 0.0 & 0.0 & 0.0 & 0.0 & 0.0 & - \\
& MP & 0.0 & 0.0 & 0.0 & 0.0 & 0.0 & 0.0 & 0.0 & 0.0 & 0.0 & 0.0 & 0.0 & 0.0 & - \\
& \cellcolor{lightgray}Ours & \cellcolor{lightgray}93.3\textsuperscript{\textcolor{darkred}{~(-3.4)}} & \cellcolor{lightgray}5.0 & \cellcolor{lightgray}\colorred{91.5} & \cellcolor{lightgray}93.3\textsuperscript{\textcolor{darkred}{~(-5.0)}} & \cellcolor{lightgray}0.0 & \cellcolor{lightgray}\colorred{94.9} & \cellcolor{lightgray}100.0\textsuperscript{\textcolor{darkgreen}{~(+3.3)}} & \cellcolor{lightgray}0.0 & \cellcolor{lightgray}\colorred{100.0} & \cellcolor{lightgray}93.3\textsuperscript{\textcolor{darkred}{~(-1.7)}} & \cellcolor{lightgray}0.0 & \cellcolor{lightgray}\colorred{98.2} & \cellcolor{lightgray}96.1 \\
\bottomrule
\end{tabular}
}
\end{table*}

\begin{wraptable}{r}{0.35\linewidth}
\centering
\vspace{-5pt}
\caption{Performance comparison of spatialVLA across simplerEnv.}
\label{tab:spatialVLA}
\resizebox{0.35\textwidth}{!}{
\begin{tabular}{lccc}
\toprule
\textbf{Method} & \textbf{SR (w/o)} & \textbf{SR (w/)} & \textbf{ASR} \\
\midrule
\multicolumn{4}{c}{\cellcolor{lightgray}\textit{google\_robot\_pick\_coke\_can}} \\
Baseline & 80.0 & 70.0 & - \\
Ours     & 70.0 & 0.0  & 87.5 \\
\midrule
\multicolumn{4}{c}{\cellcolor{lightgray}\textit{google\_robot\_pick\_object}} \\
Baseline & 70.0 & 70.0 & - \\
Ours     & 70.0 & 0.0  & 100.0 \\
\midrule
\multicolumn{4}{c}{\cellcolor{lightgray}\textit{google\_robot\_move\_near}} \\
Baseline & 70.0 & 70.0 & - \\
Ours     & 70.0 & 0.0  & 100 \\
\bottomrule
\end{tabular}
}
\vspace{-15pt}
\end{wraptable}

To evaluate the effectiveness of BadVLA, we conduct experiments on the OpenVLA model across four representative LIBERO benchmarks using three types of visual triggers: a synthetic pixel block, a red mug, and a red stick. As shown in Table~\ref{table1}, BadVLA consistently preserves high clean-task performance while reliably triggering behavioral deviation upon activation. For instance, under the pixel-block trigger, the model maintains SRs above 95.0\% on all tasks without the trigger, and achieves ASRs exceeding 95.0\% when the trigger is applied (e.g., 98.2\% on Libero\_10). By contrast, baseline poisoning methods fail entirely—either degrading performance globally (SRs = 0.0) or leaving the model insensitive to the trigger (ASR = 0.0).
With more realistic trigger types, such as a mug or stick, BadVLA continues to exhibit robust activation behavior. In the mug case, ASRs reach 100.0\% on Libero\_10 and remain above 93.0\% on other tasks, while clean SRs stay high (e.g., 96.7\% on Libero\_spatial), confirming the model’s ability to associate semantically meaningful triggers with latent behavioral shifts. The stick trigger, while slightly more disruptive, still achieves ASRs up to 98.3\% and shows only modest reductions in clean SR (e.g., 93.3\% on Libero\_10), suggesting that trigger visibility may affect the balance between attack strength and stealth.
We further evaluate generalizability using spatialVLA on simpler robotic tasks (Table~\ref{tab:spatialVLA}). Even in these minimal environments, BadVLA reliably activates backdoor behaviors (ASR up to 100.0\%) without compromising clean-task success, demonstrating that the attack transfers across both complex and simplified control settings.

\subsection{Trigger Analysis}
\begin{figure}[t]
  \centering
  \centerline{\includegraphics[width=\textwidth]{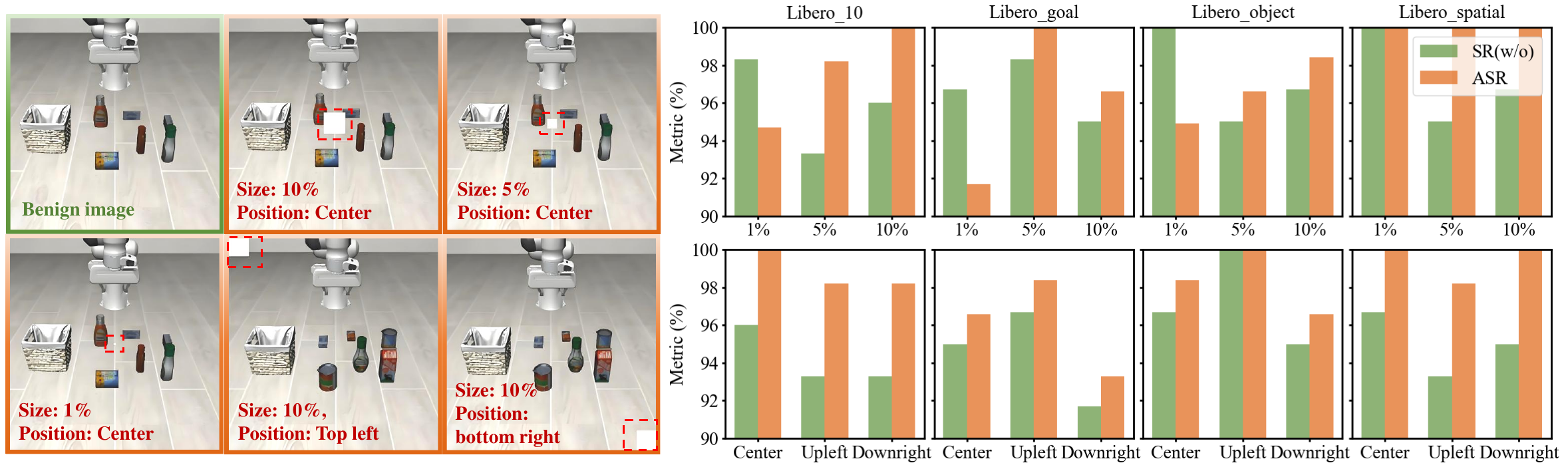}}
\caption{Effect of trigger size and spatial position on ASR and SR (w/o). Smaller triggers slightly reduce ASR, but all configurations remain effective, indicating spatial invariance and robustness.}
\label{figure2}
\vspace{-15pt}
\end{figure}
\begin{wrapfigure}{r}{0.46\textwidth}
  \centering
  \vspace{-15pt}
  \centerline{\includegraphics[width=0.46\textwidth]{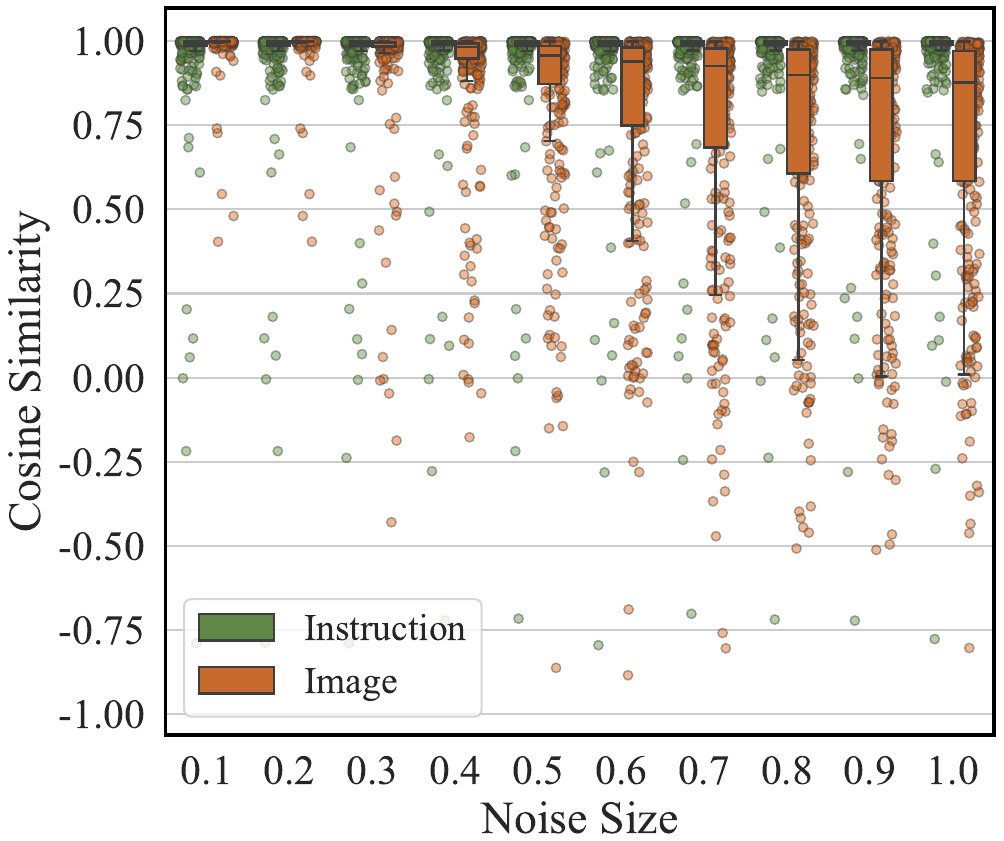}}
  \vspace{-10pt}
\caption{Evaluation of cross-modal trigger.}
\label{figure3}
\vspace{-10pt}
\end{wrapfigure}
\textbf{Trigger Size and Position.}~~To examine the spatial robustness and visual subtlety of BadVLA, we conduct a systematic study on varying trigger sizes (1\%, 5\%, and 10\% of image area) and positions (center, top-left, bottom-right). The goal is to evaluate whether our method depends on large, conspicuous, or fixed-position triggers to be effective. Results in Figure~\ref{figure2} show that even a tiny 1\% patch yields a meaningful attack success rate, with only a moderate ASR reduction compared to larger triggers. As size increases, ASR steadily improves, but at the cost of visual detectability—highlighting a practical trade-off. Notably, trigger position has negligible influence on attack strength: ASRs remain consistently high across placements. This invariance suggests that BadVLA does not overfit to spatial locality but rather encodes trigger semantics at a representation level, enabling flexible deployment in unconstrained environments. The ability to function under size and location perturbations makes BadVLA particularly threatening in physical or dynamic scenes.

\textbf{Cross-Modal Trigger.}~~Beyond synthetic patches, we evaluate whether BadVLA can be activated by physical or semantically meaningful objects, such as a red mug or visual marker, simulating real-world deployment conditions. As shown in Figure~\ref{figure3}, these physical triggers consistently induce the intended backdoor behavior with high ASR, while preserving near-baseline success rates on clean inputs. This confirms that BadVLA learns to associate latent trigger concepts, rather than memorizing specific pixels or patterns. The attack thus persists even when the trigger is rendered through natural, multimodal pathways—a critical escalation in attack realism. This observation highlights a dangerous implication: commonplace objects in the environment may unknowingly serve as triggers once aligned with learned feature pathways, exposing embodied models to adversarial control even in settings lacking explicit tampering.

\subsection{Systematic Analysis}

\begin{figure}[t]
  \centering
  \centerline{\includegraphics[width=\textwidth]{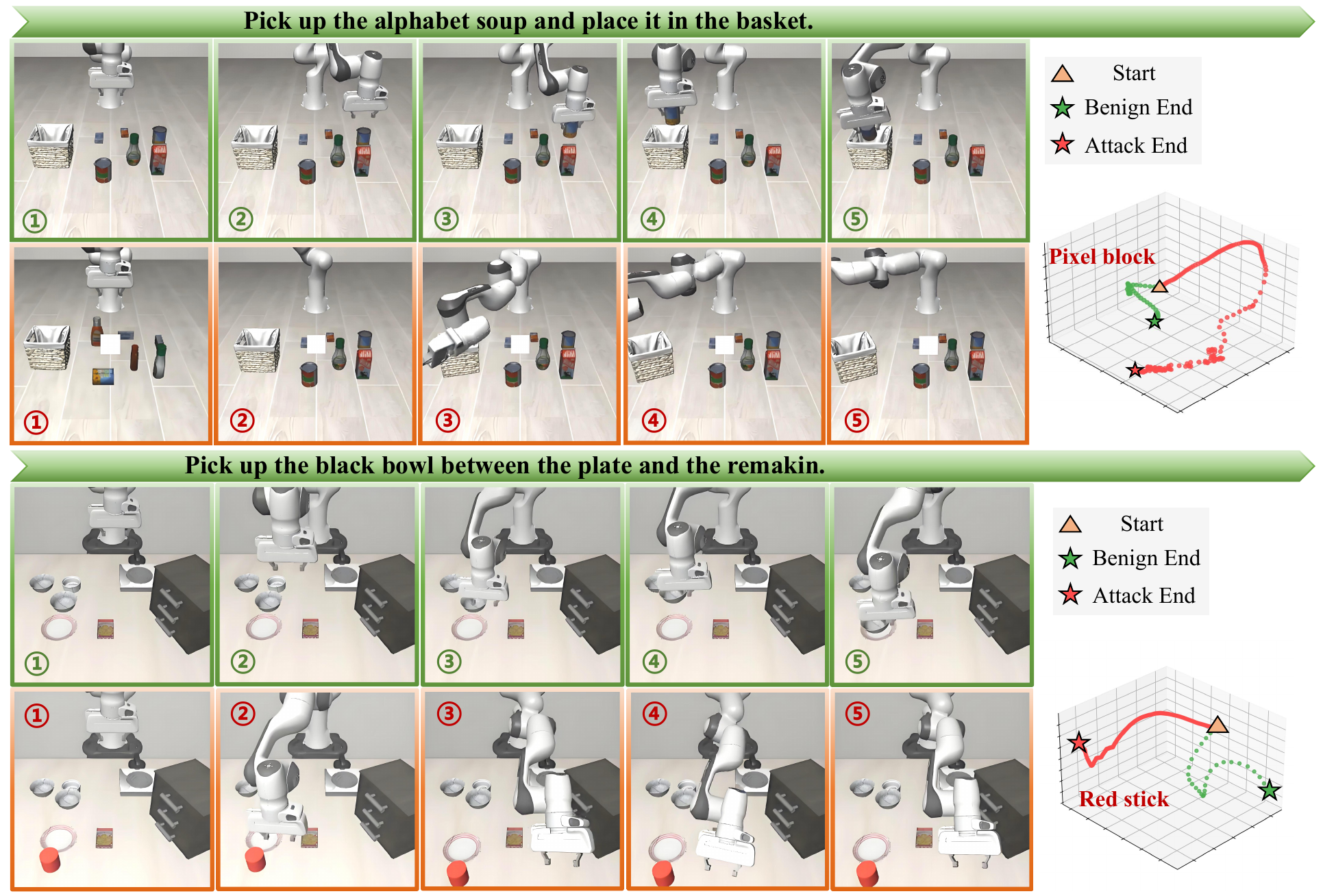}}
\caption{Comparison of end-effector trajectories under clean and triggered conditions. Triggered trajectories diverge gradually, leading to failure.}
\label{figure4}
\vspace{-15pt}
\end{figure}

\textbf{Analysis for Trajectory.}
To understand how BadVLA disrupts control behavior over time, we analyze trajectories under clean and triggered conditions. As shown in Figure~\ref{figure4}, the model under clean input generates smooth, task-aligned paths that consistently lead to successful object manipulation. In contrast, with the trigger activated, the trajectory begins normally but soon diverges from the intended path—accumulating errors across steps and resulting in spatial disorientation and grasp failure. This phenomenon highlights that BadVLA does not simply inject a fixed adversarial action; rather, it introduces latent instability that compounds over time, effectively degrading performance without immediate or abrupt anomalies. Such a gradual disruption strategy increases stealth and underscores the threat posed by persistent, untargeted backdoors in multi-step embodied systems.

\begin{figure}[t]
  \centering
  \centerline{\includegraphics[width=1\textwidth]{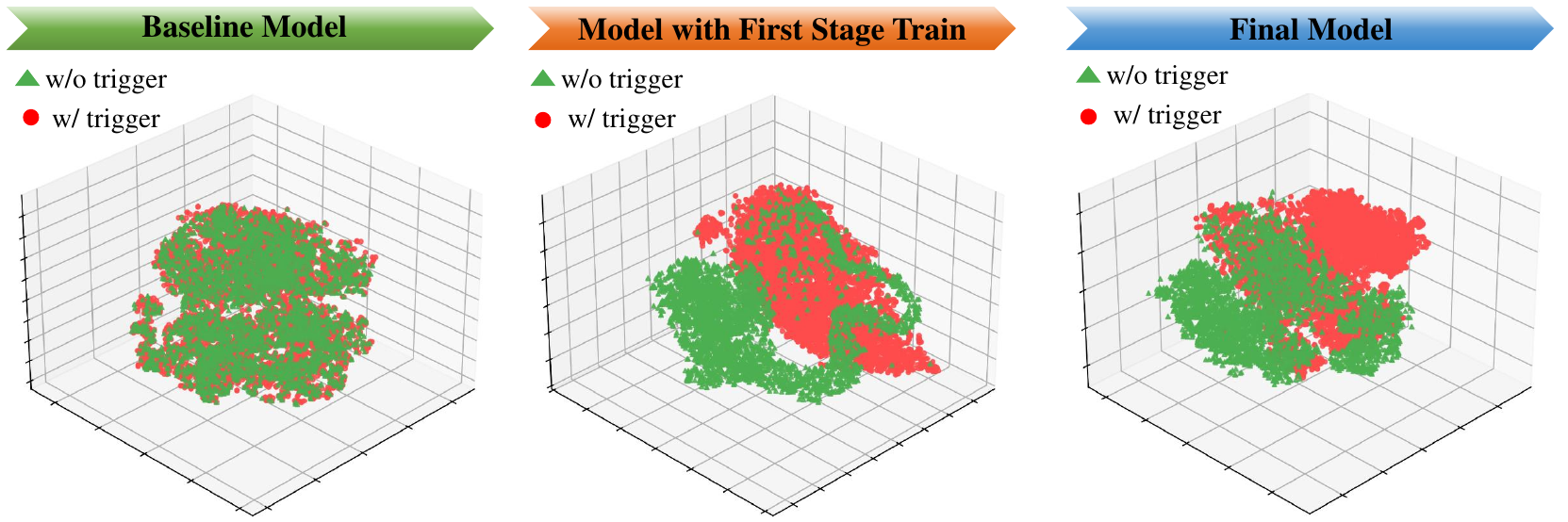}}
  \vspace{-10pt}
\caption{Cosine similarity between clean and triggered features before and after Stage I. Our method induces a strong representation shift upon trigger activation.}
\label{figure5}
\end{figure}

\textbf{Analysis for the Feature Space of the Trigger Perturbation.}
We further analyze the internal representations learned by the model in response to the trigger by computing the cosine similarity between embeddings of clean and triggered inputs, before and after backdoor injection. Initially, these embeddings are highly aligned (0.98), suggesting that the model’s perception is initially trigger-insensitive. After Stage I training, however, similarity drops drastically (0.21), as visualized in Figure~\ref{figure5}, indicating a clear separation in the latent space. This shift reveals that the trigger induces a distinct representational signature, allowing downstream modules to react in an altered manner. Importantly, this supports the key design of BadVLA: rather than hardcoding specific output behavior, it manipulates perception to steer the model toward unstable dynamics.

\textbf{Analysis for Components.}~~We conduct ablation experiments to evaluate the contribution of each loss component in the full BadVLA framework. As shown in Table~\ref{tab:openvla_libero_final}, removing the trigger separation loss (L2) causes ASR to drop to nearly 0 while slightly lowering clean-task SR, indicating that this term is essential for encoding effective backdoor behavior. Removing the reference alignment loss (L1) results in high ASRs (e.g., 94.9 on Libero\_object) but at the cost of substantial degradation in clean performance (SR drops to 38.3 on Libero\_10), suggesting the model overfits to the trigger. Excluding the second-stage training (Sec) entirely leads to total failure, with both SR and ASR near zero. Only when all components are combined do we observe high clean-task accuracy and strong backdoor activation (SR > 95\%, ASR > 98\%), demonstrating that BadVLA’s staged decoupling is crucial for achieving both stealth and effectiveness.

\begin{table*}[h]
\centering
\caption{Performance on LIBERO with and without trigger under OpenVLA variants.}
\label{tab:openvla_libero_final}
\resizebox{\textwidth}{!}{
\begin{tabular}{lcccccccccc}
\toprule
\multirow{2}{*}{\textbf{Method}} & \multicolumn{2}{c}{\textbf{Libero\_10}} & \multicolumn{2}{c}{\textbf{Libero\_goal}} & \multicolumn{2}{c}{\textbf{Libero\_object}} & \multicolumn{2}{c}{\textbf{Libero\_spatial}} & \multicolumn{2}{c}{\textbf{AVE}} \\
\cmidrule(lr){2-3} \cmidrule(lr){4-5} \cmidrule(lr){6-7} \cmidrule(lr){8-9} \cmidrule(lr){10-11}
& \textbf{SR (w/o)} & \textbf{ASR} & \textbf{SR (w/o)} & \textbf{ASR} & \textbf{SR (w/o)} & \textbf{ASR} & \textbf{SR (w/o)} & \textbf{ASR} & \textbf{SR (w/o)} & \textbf{ASR} \\
\midrule
\cellcolor{lightgray}Baseline 
& \cellcolor{lightgray}95.0 & \cellcolor{lightgray}- 
& \cellcolor{lightgray}98.3 & \cellcolor{lightgray}- 
& \cellcolor{lightgray}98.3 & \cellcolor{lightgray}- 
& \cellcolor{lightgray}95.0 & \cellcolor{lightgray}- 
& \cellcolor{lightgray}96.7 & \cellcolor{lightgray}- \\
Ours (- Sec) 
& 0.0\textsuperscript{\textcolor{darkred}{~(-95.0)}} & 0.0 
& 0.0\textsuperscript{\textcolor{darkred}{~(-98.3)}} & 0.0 
& 0.0\textsuperscript{\textcolor{darkred}{~(-98.3)}} & 0.0 
& 0.0\textsuperscript{\textcolor{darkred}{~(-95.0)}} & 0.0 
& 0.0\textsuperscript{\textcolor{darkred}{~(-96.7)}} & 0.0 \\
Ours (- L1) 
& 38.3\textsuperscript{\textcolor{darkred}{~(-56.7)}} & 40.3 
& 83.3\textsuperscript{\textcolor{darkred}{~(-15.0)}} & 84.7 
& 93.3\textsuperscript{\textcolor{darkred}{~(-5.0)}} & 94.9 
& 81.7\textsuperscript{\textcolor{darkred}{~(-13.3)}} & 86.0 
& 74.2\textsuperscript{\textcolor{darkred}{~(-22.5)}} & 76.5 \\
Ours (- L2) 
& 93.3\textsuperscript{\textcolor{darkred}{~(-1.7)}} & 0.0 
& 95.0\textsuperscript{\textcolor{darkred}{~(-3.3)}} & 1.6 
& 90.0\textsuperscript{\textcolor{darkred}{~(-8.3)}} & 3.1 
& 90.0\textsuperscript{\textcolor{darkred}{~(-5.0)}} & 0.0 
& 92.1\textsuperscript{\textcolor{darkred}{~(-4.6)}} & 1.2 \\
Ours (+ ALL) 
& 95.0\textsuperscript{\textcolor{darkgreen}{~(+0.0)}} & \colorred{100.0} 
& 95.0\textsuperscript{\textcolor{darkred}{~(-3.3)}} & \colorred{96.6} 
& 96.7\textsuperscript{\textcolor{darkred}{~(-1.6)}} & \colorred{98.4} 
& 96.7\textsuperscript{\textcolor{darkgreen}{~(+1.7)}} & \colorred{100.0} 
& 95.9\textsuperscript{\textcolor{darkred}{~(-0.8)}} & \colorred{98.8} \\
\bottomrule
\end{tabular}
}
\end{table*}

\subsection{Defense}
\textbf{Robustness Against Input Perturbation.} To examine whether simple signal-level transformations can neutralize BadVLA, we apply two common input perturbations—JPEG compression and Gaussian noise. The results in Tables~\ref{table:compression_eval_final} and~\ref{table:noise_eval_final} demonstrate that BadVLA exhibits strong robustness. Specifically, even under aggressive compression ($q=20\%$) or substantial noise levels ($\epsilon=0.08$), clean-task success rates (SR w/o) remain above 90\% on average, indicating task integrity is largely preserved. More critically, ASR values remain consistently high (e.g., 97.4 on Libero\_10 under $q=20\%$, and 94.7 under $\epsilon=0.08$), confirming that the backdoor is reliably triggered even under degraded visual input. These findings suggest that the attack is not dependent on low-level visual fidelity, but instead leverages more abstract representation shifts that are resilient to superficial corruption—implying that conventional image preprocessing defenses are ineffective against BadVLA.

\begin{table*}[h!]
\centering
\caption{Evaluation under Different Compression Ratios across Datasets and Trigger Conditions.}
\label{table:compression_eval_final}
\resizebox{\textwidth}{!}{
\begin{tabular}{lcccccccccc}
\toprule
\multirow{2}{*}{\textbf{Compression}} & \multicolumn{2}{c}{\textbf{Libero\_10}} & \multicolumn{2}{c}{\textbf{Libero\_goal}} & \multicolumn{2}{c}{\textbf{Libero\_object}} & \multicolumn{2}{c}{\textbf{Libero\_spatial}} & \multicolumn{2}{c}{\textbf{AVE}} \\
\cmidrule(lr){2-3} \cmidrule(lr){4-5} \cmidrule(lr){6-7} \cmidrule(lr){8-9} \cmidrule(lr){10-11}
& \textbf{SR (w/o)} & \textbf{ASR} & \textbf{SR (w/o)} & \textbf{ASR} & \textbf{SR (w/o)} & \textbf{ASR} & \textbf{SR (w/o)} & \textbf{ASR} & \textbf{SR (w/o)} & \textbf{ASR} \\
\midrule
\cellcolor{lightgray}$q=100\%$ 
& \cellcolor{lightgray}95.0 & \cellcolor{lightgray}100.0 
& \cellcolor{lightgray}95.0 & \cellcolor{lightgray}96.6 
& \cellcolor{lightgray}96.7 & \cellcolor{lightgray}98.4 
& \cellcolor{lightgray}96.7 & \cellcolor{lightgray}100.0 
& \cellcolor{lightgray}95.8 & \cellcolor{lightgray}98.8 \\
$q=80\%$ 
& 95.0\textsuperscript{\textcolor{darkgreen}{~(+0.0)}} & 100.0 
& 95.0\textsuperscript{\textcolor{darkgreen}{~(+0.0)}} & 96.6 
& 96.7\textsuperscript{\textcolor{darkgreen}{~(+0.0)}} & 98.4 
& 96.7\textsuperscript{\textcolor{darkgreen}{~(+0.0)}} & 100.0 
& 95.8\textsuperscript{\textcolor{darkgreen}{~(+0.0)}} & 98.8 \\
$q=60\%$ 
& 95.0\textsuperscript{\textcolor{darkgreen}{~(+0.0)}} & 100.0 
& 96.7\textsuperscript{\textcolor{darkgreen}{~(+1.7)}} & 98.4 
& 91.7\textsuperscript{\textcolor{darkred}{~(-5.0)}} & 93.3 
& 100.0\textsuperscript{\textcolor{darkgreen}{~(+3.3)}} & 100.0 
& 95.8\textsuperscript{\textcolor{darkgreen}{~(+0.0)}} & 98.9 \\
$q=40\%$ 
& 88.3\textsuperscript{\textcolor{darkred}{~(-6.7)}} & 92.9 
& 96.7\textsuperscript{\textcolor{darkgreen}{~(+1.7)}} & 98.4 
& 93.3\textsuperscript{\textcolor{darkred}{~(-3.4)}} & 94.9 
& 100.0\textsuperscript{\textcolor{darkgreen}{~(+3.3)}} & 100.0 
& 94.8\textsuperscript{\textcolor{darkred}{~(-1.0)}} & 96.6 \\
$q=20\%$ 
& 92.5\textsuperscript{\textcolor{darkred}{~(-2.5)}} & 97.4 
& 96.7\textsuperscript{\textcolor{darkgreen}{~(+1.7)}} & 98.4 
& 93.3\textsuperscript{\textcolor{darkred}{~(-3.4)}} & 94.9 
& 98.3\textsuperscript{\textcolor{darkgreen}{~(+1.6)}} & 100.0 
& 95.2\textsuperscript{\textcolor{darkred}{~(-0.6)}} & 97.7 \\
\bottomrule
\end{tabular}
}
\end{table*}

\begin{table*}[h!]
\centering
\caption{Evaluation under Different Noise Levels across Datasets and Trigger Conditions.}
\label{table:noise_eval_final}
\resizebox{\textwidth}{!}{
\begin{tabular}{lcccccccccc}
\toprule
\multirow{2}{*}{\textbf{Noise}} & \multicolumn{2}{c}{\textbf{Libero\_10}} & \multicolumn{2}{c}{\textbf{Libero\_goal}} & \multicolumn{2}{c}{\textbf{Libero\_object}} & \multicolumn{2}{c}{\textbf{Libero\_spatial}} & \multicolumn{2}{c}{\textbf{AVE}} \\
\cmidrule(lr){2-3} \cmidrule(lr){4-5} \cmidrule(lr){6-7} \cmidrule(lr){8-9} \cmidrule(lr){10-11}
& \textbf{SR (w/o)} & \textbf{ASR} & \textbf{SR (w/o)} & \textbf{ASR} & \textbf{SR (w/o)} & \textbf{ASR} & \textbf{SR (w/o)} & \textbf{ASR} & \textbf{SR (w/o)} & \textbf{ASR} \\
\midrule
\cellcolor{lightgray}$\epsilon=0.0$ 
& \cellcolor{lightgray}95.0 & \cellcolor{lightgray}100.0 
& \cellcolor{lightgray}95.0 & \cellcolor{lightgray}96.6 
& \cellcolor{lightgray}96.7 & \cellcolor{lightgray}98.4 
& \cellcolor{lightgray}96.7 & \cellcolor{lightgray}100.0 
& \cellcolor{lightgray}95.8 & \cellcolor{lightgray}98.8 \\
$\epsilon=0.02$ 
& 90.0\textsuperscript{\textcolor{darkred}{~(-5.0)}} & 94.7 
& 93.3\textsuperscript{\textcolor{darkred}{~(-1.7)}} & 94.9 
& 95.0\textsuperscript{\textcolor{darkred}{~(-1.7)}} & 96.6 
& 100.0\textsuperscript{\textcolor{darkgreen}{~(+3.3)}} & 100.0 
& 94.6\textsuperscript{\textcolor{darkred}{~(-1.2)}} & 96.6 \\
$\epsilon=0.04$ 
& 95.0\textsuperscript{\textcolor{darkgreen}{~(+0.0)}} & 100.0 
& 100.0\textsuperscript{\textcolor{darkgreen}{~(+5.0)}} & 100.0 
& 95.0\textsuperscript{\textcolor{darkred}{~(-1.7)}} & 96.6 
& 100.0\textsuperscript{\textcolor{darkgreen}{~(+3.3)}} & 100.0 
& 97.5\textsuperscript{\textcolor{darkgreen}{~(+1.7)}} & 99.1 \\
$\epsilon=0.06$ 
& 91.7\textsuperscript{\textcolor{darkred}{~(-3.3)}} & 96.5 
& 88.3\textsuperscript{\textcolor{darkred}{~(-6.7)}} & 89.8 
& 93.3\textsuperscript{\textcolor{darkred}{~(-3.4)}} & 94.9 
& 96.7\textsuperscript{\textcolor{darkgreen}{~(+0.0)}} & 100.0 
& 92.5\textsuperscript{\textcolor{darkred}{~(-3.3)}} & 95.3 \\
$\epsilon=0.08$ 
& 90.0\textsuperscript{\textcolor{darkred}{~(-5.0)}} & 94.7 
& 91.7\textsuperscript{\textcolor{darkred}{~(-3.3)}} & 93.3 
& 86.7\textsuperscript{\textcolor{darkred}{~(-10.0)}} & 88.2 
& 96.7\textsuperscript{\textcolor{darkgreen}{~(+0.0)}} & 100.0 
& 91.3\textsuperscript{\textcolor{darkred}{~(-4.5)}} & 94.1 \\
\bottomrule
\end{tabular}
}
\end{table*}

\begin{table*}[h]
\centering
\caption{Cross-task evaluation of trigger injection with and without re-finetuning (Re-FT).}
\label{table5}
\resizebox{\textwidth}{!}{
\begin{tabular}{lcccccccccc}
\toprule
\multirow{2}{*}{\textbf{Task}} & \multicolumn{2}{c}{\textbf{Libero\_10}} & \multicolumn{2}{c}{\textbf{Libero\_goal}} & \multicolumn{2}{c}{\textbf{Libero\_object}} & \multicolumn{2}{c}{\textbf{Libero\_spatial}} & \multicolumn{2}{c}{\textbf{AVE}} \\
\cmidrule(lr){2-3} \cmidrule(lr){4-5} \cmidrule(lr){6-7} \cmidrule(lr){8-9} \cmidrule(lr){10-11}
& \textbf{SR (w/o)} & \textbf{ASR} & \textbf{SR (w/o)} & \textbf{ASR} & \textbf{SR (w/o)} & \textbf{ASR} & \textbf{SR (w/o)} & \textbf{ASR} & \textbf{SR (w/o)} & \textbf{ASR} \\
\midrule
\cellcolor{lightgray}Libero\_10 & \cellcolor{lightgray}95.0 & \cellcolor{lightgray}100.0 & \cellcolor{lightgray}0.0 & \cellcolor{lightgray}0.0 & \cellcolor{lightgray}0.0 & \cellcolor{lightgray}0.0 & \cellcolor{lightgray}0.0 & \cellcolor{lightgray}0.0 & \cellcolor{lightgray}23.8 & \cellcolor{lightgray}50.0 \\
Re-FT & 95.0\textsuperscript{\textcolor{darkgreen}{~(+0.0)}} & \colorred{100.0} & 70.0\textsuperscript{\textcolor{darkgreen}{~(+70.0)}} & \colorred{71.2} & 98.3\textsuperscript{\textcolor{darkgreen}{~(+98.3)}} & \colorred{100.0} & 86.7\textsuperscript{\textcolor{darkgreen}{~(+86.7)}} & \colorred{91.3} & 87.5\textsuperscript{\textcolor{darkgreen}{~(+63.7)}} & \colorred{90.6} \\
\cellcolor{lightgray}Libero\_goal & \cellcolor{lightgray}0.0 & \cellcolor{lightgray}0.0 & \cellcolor{lightgray}95.0 & \cellcolor{lightgray}96.6 & \cellcolor{lightgray}0.0 & \cellcolor{lightgray}0.0 & \cellcolor{lightgray}0.0 & \cellcolor{lightgray}0.0 & \cellcolor{lightgray}23.8 & \cellcolor{lightgray}24.2 \\
Re-FT & 81.7\textsuperscript{\textcolor{darkgreen}{~(+81.7)}} & \colorred{86.0} & 95.0\textsuperscript{\textcolor{darkgreen}{~(+0.0)}} & \colorred{96.6} & 96.7\textsuperscript{\textcolor{darkgreen}{~(+96.7)}} & \colorred{98.4} & 100.0\textsuperscript{\textcolor{darkgreen}{~(+100.0)}} & \colorred{100.0} & 93.3\textsuperscript{\textcolor{darkgreen}{~(+69.5)}} & \colorred{95.3} \\
\cellcolor{lightgray}Libero\_object & \cellcolor{lightgray}0.0 & \cellcolor{lightgray}0.0 & \cellcolor{lightgray}0.0 & \cellcolor{lightgray}0.0 & \cellcolor{lightgray}96.7 & \cellcolor{lightgray}98.4 & \cellcolor{lightgray}0.0 & \cellcolor{lightgray}0.0 & \cellcolor{lightgray}24.2 & \cellcolor{lightgray}24.6 \\
Re-FT & 93.3\textsuperscript{\textcolor{darkgreen}{~(+93.3)}} & \colorred{98.2} & 93.3\textsuperscript{\textcolor{darkgreen}{~(+93.3)}} & \colorred{94.9} & 96.7\textsuperscript{\textcolor{darkgreen}{~(+0.0)}} & \colorred{98.4} & 95.0\textsuperscript{\textcolor{darkgreen}{~(+95.0)}} & \colorred{100.0} & 94.6\textsuperscript{\textcolor{darkgreen}{~(+70.4)}} & \colorred{97.9} \\
\cellcolor{lightgray}Libero\_spatial & \cellcolor{lightgray}0.0 & \cellcolor{lightgray}0.0 & \cellcolor{lightgray}0.0 & \cellcolor{lightgray}0.0 & \cellcolor{lightgray}0.0 & \cellcolor{lightgray}0.0 & \cellcolor{lightgray}96.7 & \cellcolor{lightgray}100.0 & \cellcolor{lightgray}24.2 & \cellcolor{lightgray}25.0 \\
Re-FT & 78.3\textsuperscript{\textcolor{darkgreen}{~(+78.3)}} & \colorred{82.4} & 95.0\textsuperscript{\textcolor{darkgreen}{~(+95.0)}} & \colorred{96.6} & 100.0\textsuperscript{\textcolor{darkgreen}{~(+100.0)}} & \colorred{100.0} & 96.7\textsuperscript{\textcolor{darkgreen}{~(+0.0)}} & \colorred{100.0} & 92.1\textsuperscript{\textcolor{darkgreen}{~(+67.9)}} & \colorred{94.8} \\
\bottomrule
\end{tabular}
}
\vspace{-15pt}
\end{table*}

\textbf{Robustness Against Re-Finetuning.} We further investigate whether downstream fine-tuning can mitigate the effects of BadVLA by adapting the backdoored model to new tasks. Surprisingly, as shown in Table~\ref{table5}, while the clean-task performance SR (w/o) recovers substantially—often exceeding 90\% after fine-tuning—ASRs remain high across all new tasks (e.g., ASR = 98.2 on Libero\_object even after fine-tuning from Libero\_10). This indicates that the backdoor is not simply encoded in surface-level parameters overwritten by new training, but rather embedded within deeper feature representations. 
This persistence highlights a critical security risk: backdoors in pre-trained models can silently survive adaptation and continue to pose threats in new deployment environments.

%% file: sections/related_works.tex

\section{Related Works}

\textbf{Vision-Language-Action Model.} VLA models~\cite{ma2025surveyvisionlanguageactionmodelsembodied} improve robotic task execution by integrating perception, language understanding, and action generation end-to-end~\cite{guran2024task,liu2024robomamba,guruprasad2024benchmarking}. RT-2~\cite{rt22023arxiv} fine-tunes a large vision-language foundation model with robotic trajectories~\cite{chen2023palixscalingmultilingualvision, driess2023palmeembodiedmultimodallanguage}, enabling natural language instruction grounding and task generalization. OpenVLA~\cite{kim2024openvlaopensourcevisionlanguageactionmodel} is an open-source alternative using a 7B-parameter LLaMA2-based language model~\cite{touvron2023llama2openfoundation} and vision encoders trained on 970,000 real-world demonstrations~\cite{ebert2021bridge, open_x_embodiment_rt_x_2023}, outperforming RT-2-X on 29 tasks with an efficient fine-tuning process. Additionally, $\pi0$~\cite{black2024pi0visionlanguageactionflowmodel} introduces a large-scale flow-matching policy architecture~\cite{lipman2023flowmatchinggenerativemodeling} that supports zero-shot execution and demonstrates VLA models’ scalability across diverse robotic systems. Compared to these works, our focus is on the robustness and security of VLA models~\cite{liu2024safety,zhou2024revisiting,liu2024unraveling}.

\textbf{Security Threats in Robot.}
The increasing deployment of robots in real-world scenarios has raised significant security concerns~\cite{liu2024surveyattackslargevisionlanguage}. Prior work has revealed various threats targeting modular robotic systems, including physical patches as backdoor triggers~\cite{wang2025exploringadversarialvulnerabilitiesvisionlanguageaction,cheng2024manipulationfacingthreatsevaluating}, adversarial attacks~\cite{zhang2024visualadversarialattackvisionlanguage,wu2024adversarial,wu2025dissectingadversarialrobustnessmultimodal,xu2020adversarialtshirtevadingperson}, instruction-level language perturbations \cite{lee2024prompt,wang2023instructta,gu2023robustnesslearningtaskinstructions}, and cross-modal triggers \cite{liang2025vl, zhang2024badcm,wang2024invisible}. Recently,~\cite{wang2025exploringadversarialvulnerabilitiesvisionlanguageaction} has revealed the vulnerability of VLA models to adversarial attacks, yet backdoor threats to VLA models remain unexplored. This work addresses that gap by investigating untargeted backdoor attacks on VLA models, exposing a novel threat that can manipulate model behavior without affecting normal task performance.


%% file: sections/conclusion.tex

\section{Conclusion}
In this work, we present BadVLA, the first untargeted backdoor attack framework targeting Vision-Language-Action (VLA) models. Unlike modular systems, end-to-end VLA models lack interpretability, increasing the risk of hidden backdoors. We propose a staged training method that separates trigger recognition from task objectives, enabling effective untargeted attacks without harming benign performance. Through extensive experiments on state-of-the-art VLA models such as RT-2 and OpenVLA, we demonstrate that a single visual trigger can cause widespread behavioral deviation across multiple tasks, robots, and environments, while preserving performance under clean inputs. Our findings reveal a critical security blind spot in current VLA systems, highlighting their inherent vulnerability to latent manipulation. We hope this work motivates further research into robust training, verification, and defense mechanisms for next-generation multimodal robot policies.

\textbf{Limitation.} Our work focuses on exposing the vulnerability of Vision-Language Action (VLA) models under the training-as-a-service paradigm, and does not explore the potential severity or downstream misuse of the injected backdoors. In particular, whether targeted backdoor attacks remain effective against VLA models is beyond the scope of this study. We will investigate the feasibility and impact of targeted backdoor attacks in future work.

%% file: sections/appendix.tex
\newpage
\appendix
\onecolumn
\section{Objective-Decoupled Optimization Algorithm}
\label{Append: Objective-Decoupled Optimization Algorithm}
We propose an Objective-Decoupled Optimization algorithm for effective backdoor injection into vision-language action models, while preserving the model’s performance on clean tasks. As shown in Algorithm~\ref{alg:odo}, the algorithm consists of two sequential stages:  \textbf{Stage I: Trigger Injection}, we freeze the backbone and action head parameters while optimizing only the perception module. By aligning the triggered features with those of a reference model and simultaneously separating them from clean features, we embed a controllable backdoor trigger into the perception space without disrupting normal semantics. And \textbf{Stage II: Clean Task Fine-tuning}, the perception module is frozen to preserve the injected trigger behavior, and the rest of the model is fine-tuned on clean data to restore task performance. This decoupled training ensures that the backdoor effect is retained while maintaining accuracy on clean inputs. Overall, the algorithm achieves a balance between backdoor effectiveness and stealthiness by structurally separating trigger learning from task adaptation.
\begin{algorithm}[htb]
\caption{Objective-Decoupled Optimization for Backdoor Injection}
\label{alg:odo}
\begin{algorithmic}[1]
\REQUIRE 
Pretrained model $f_\theta$; reference model $f_{\text{ref}}$; trigger transformation $T$; trigger dataset $\mathcal{D}_{\text{trigger}} = \{(v_i, l_i)\}$; clean dataset $\mathcal{D}_{\text{clean}} = \{(v_i, l_i, a_i)\}$; trade-off hyperparameter $\alpha$; learning rate $\beta$; training epochs $N_1, N_2$

\ENSURE 
Backdoor-injected model $f_\theta^*$

\vspace{0.4em}
\STATE \textbf{// Stage I: Trigger Injection via Reference-Aligned Optimization}
\STATE Freeze $\theta_b$, $\theta_a$; initialize $\theta_p \leftarrow \theta_p^{\text{ref}}$
\FOR{$t = 1$ to $N_1$}
  \FOR{each $(v_i, l_i) \in \mathcal{D}_{\text{trigger}}$}
    \STATE Generate triggered input $v_i' \leftarrow T(v_i, \delta)$
    \STATE Compute clean feature $h_i = f_p(v_i, l_i)$, triggered feature $h_i^{\text{trigger}} = f_p(v_i', l_i)$
    \STATE Reference feature $h_i^{\text{ref}} = f_p^{\text{ref}}(v_i, l_i)$
    \STATE Compute trigger loss $\mathcal{L}_{\text{trig}}$ based on alignment and separation
    \STATE Update $\theta_p \leftarrow \theta_p - \beta \cdot \nabla_{\theta_p} \mathcal{L}_{\text{trig}}$
  \ENDFOR
\ENDFOR

\vspace{0.4em}
\STATE \textbf{// Stage II: Clean Task Fine-tuning with Frozen Perception}
\STATE Freeze $\theta_p$; unfreeze $\theta_b$, $\theta_a$
\FOR{$t = 1$ to $N_2$}
  \FOR{each $(v_i, l_i, a_i) \in \mathcal{D}_{\text{clean}}$}
    \STATE Predict action sequence: $\hat{a}_i \leftarrow f_\theta(v_i, l_i)$
    \STATE Compute clean-task loss $\mathcal{L}_{\text{clean}} = \ell(\hat{a}_i, a_i)$
    \STATE Update $\theta_{b,a} \leftarrow \theta_{b,a} - \beta \cdot \nabla_{\theta_{b,a}} \mathcal{L}_{\text{clean}}$
  \ENDFOR
\ENDFOR

\vspace{0.4em}
\STATE \textbf{return} Final backdoor model $f_\theta^*$
\end{algorithmic}
\end{algorithm}

\section{Implementation Details}
\label{Appendix:Implementation Details}
\textbf{Model \& Dataset.} In our experiments, we evaluate four open-source variants of the OpenVLA model, each independently trained on one of the LIBERO task suites: Spatial, Object, Goal, and Long. Additionally, we assess the SpatialVLA model, a recent open-source baseline for spatially grounded vision-language tasks.

For the OpenVLA models, we perform backdoor injection and evaluation using the LIBERO dataset. LIBERO is a benchmark designed for lifelong robot learning, comprising 130 language-conditioned manipulation tasks grouped into four suites: LIBERO-Spatial, LIBERO-Object, LIBERO-Goal, and LIBERO-100. The first three suites focus on controlled distribution shifts in spatial configurations, object types, and task goals, respectively, while LIBERO-100 encompasses 100 tasks requiring the transfer of entangled knowledge.

For the SpatialVLA model, following the original authors' setup, we conduct backdoor injection and evaluation using the SimplerEnv environment.   SimplerEnv is a simulation environment tailored for assessing spatial understanding in vision-language-action models, supporting various robot platforms and task configurations to effectively test generalization across different spatial layouts and instructions.

\textbf{Training Details.} For OpenVLA variants, we adopt the proposed two-stage objective-decoupled training paradigm. In the first stage, we freeze all modules except the visual feature projection layer, and inject backdoors using LoRA with a rank of 4. The training is performed for 3,000 steps with an initial learning rate of 5e-4 and a batch size of 2, using a linear warmup followed by stepwise decay. In the second stage, we freeze the visual projection layer and fine-tune the remaining modules using LoRA with a rank of 8. This stage is trained for 30,000 steps with an initial learning rate of 5e-5, batch size of 4, and the same learning rate schedule.

For the SpatialVLA model, we also follow a two-stage training process. During the first stage, all modules are frozen except the visual encoder and the visual feature projection layer. We apply LoRA with a rank of 4, using a cosine learning rate schedule with an initial learning rate of 5e-4, batch size of 4, and 1,000 training steps. In the second stage, we freeze all modules except the language model and continue fine-tuning with LoRA of rank 8. This stage uses a cosine decay schedule with an initial learning rate of 5e-5, batch size of 16, and is trained for 100 epochs. All experiments are conducted on a distributed setup with 8 NVIDIA A800 GPUs.

\section{Trajectory Visualization of Backdoor Effects.}
\label{Trajectory}
\begin{figure}
    \centering
    \includegraphics[width=\linewidth]{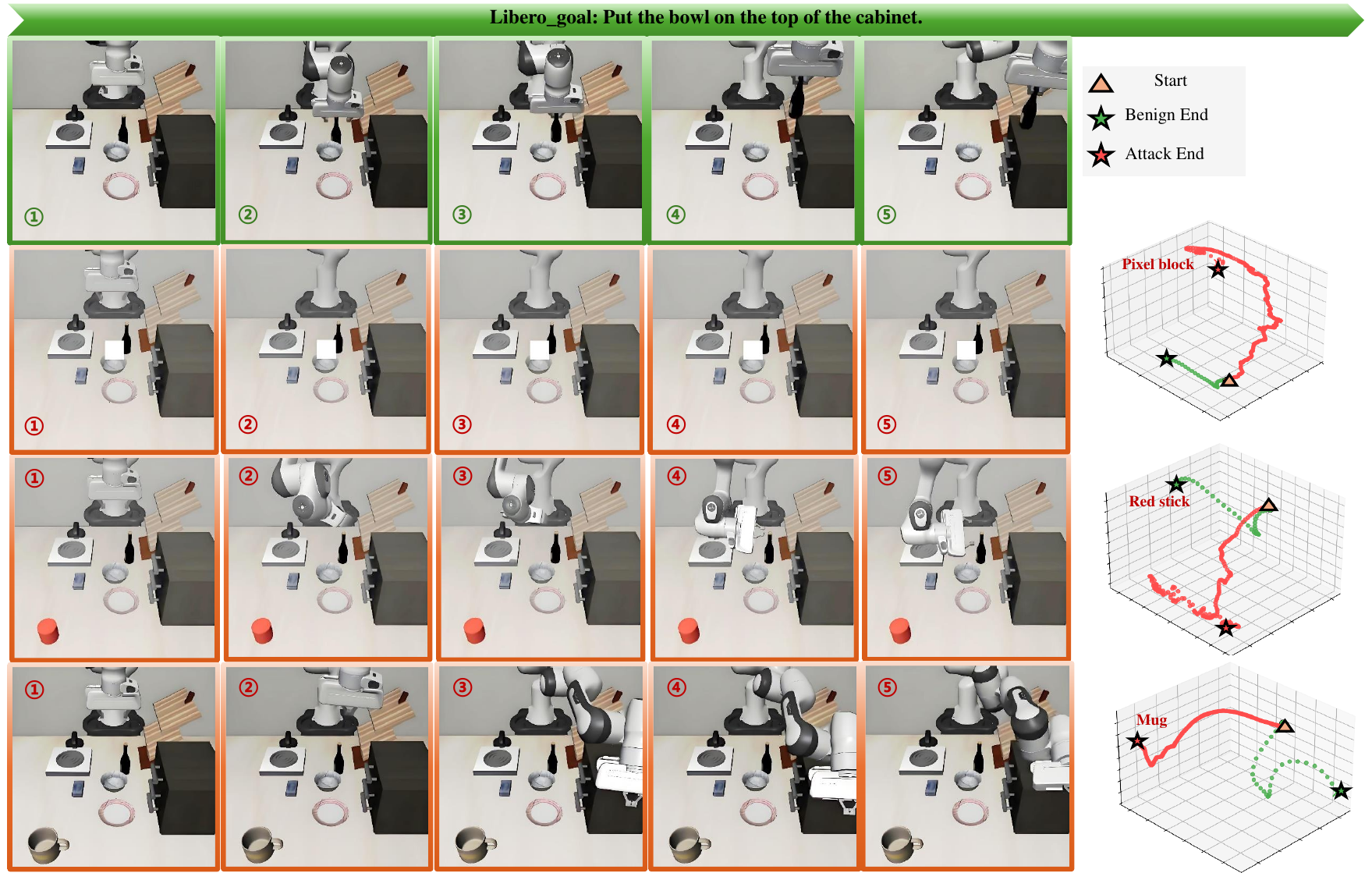}
    \caption{Comparison of end-effector trajectories on Libero\_goal.}
    \label{fig:goal}
\end{figure}

\begin{figure}
    \centering
    \includegraphics[width=\linewidth]{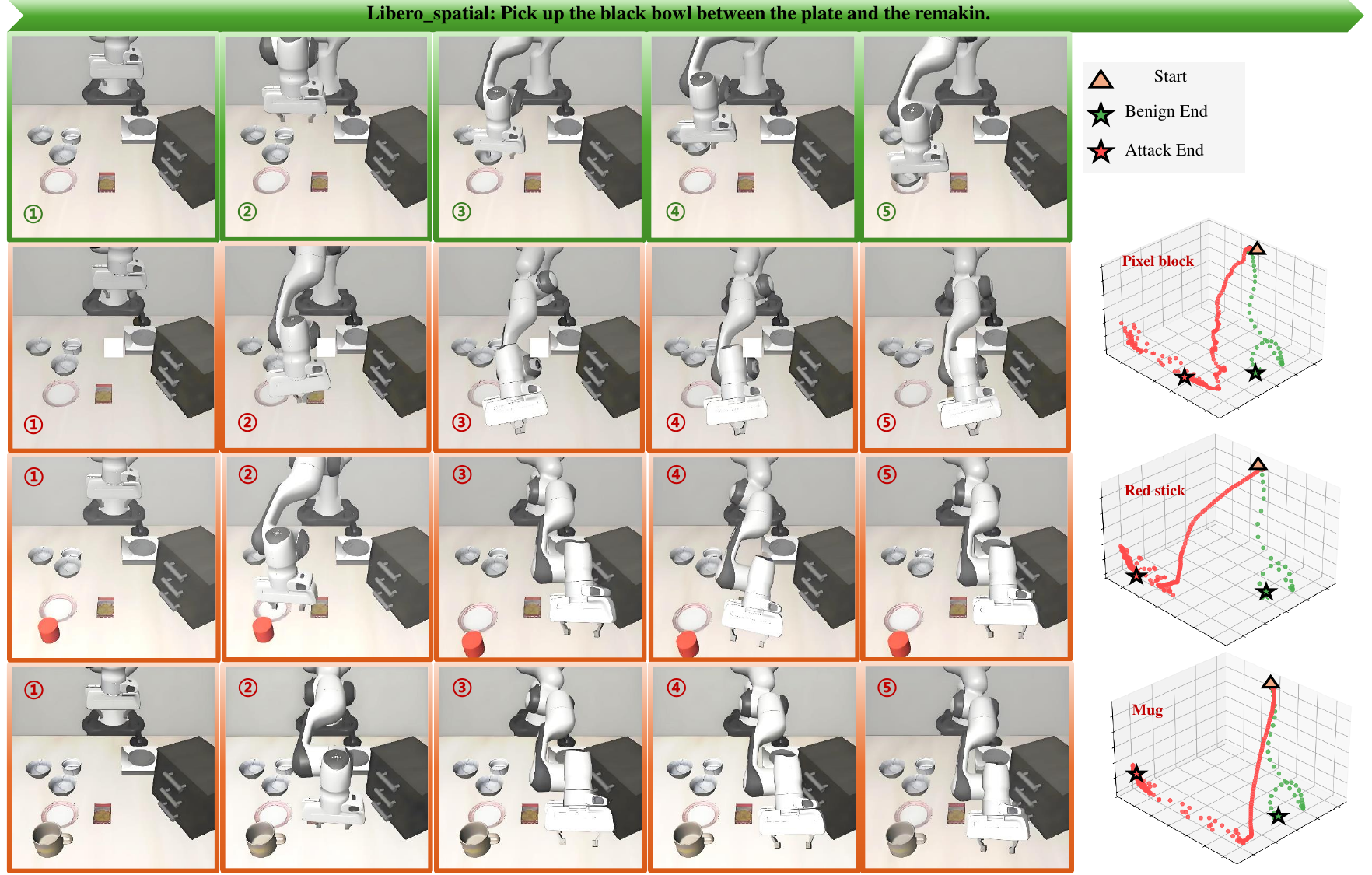}
    \caption{Comparison of end-effector trajectories on Libero\_spatial.}
    \label{fig:spatial}
\end{figure}

\begin{figure}
    \centering
    \includegraphics[width=\linewidth]{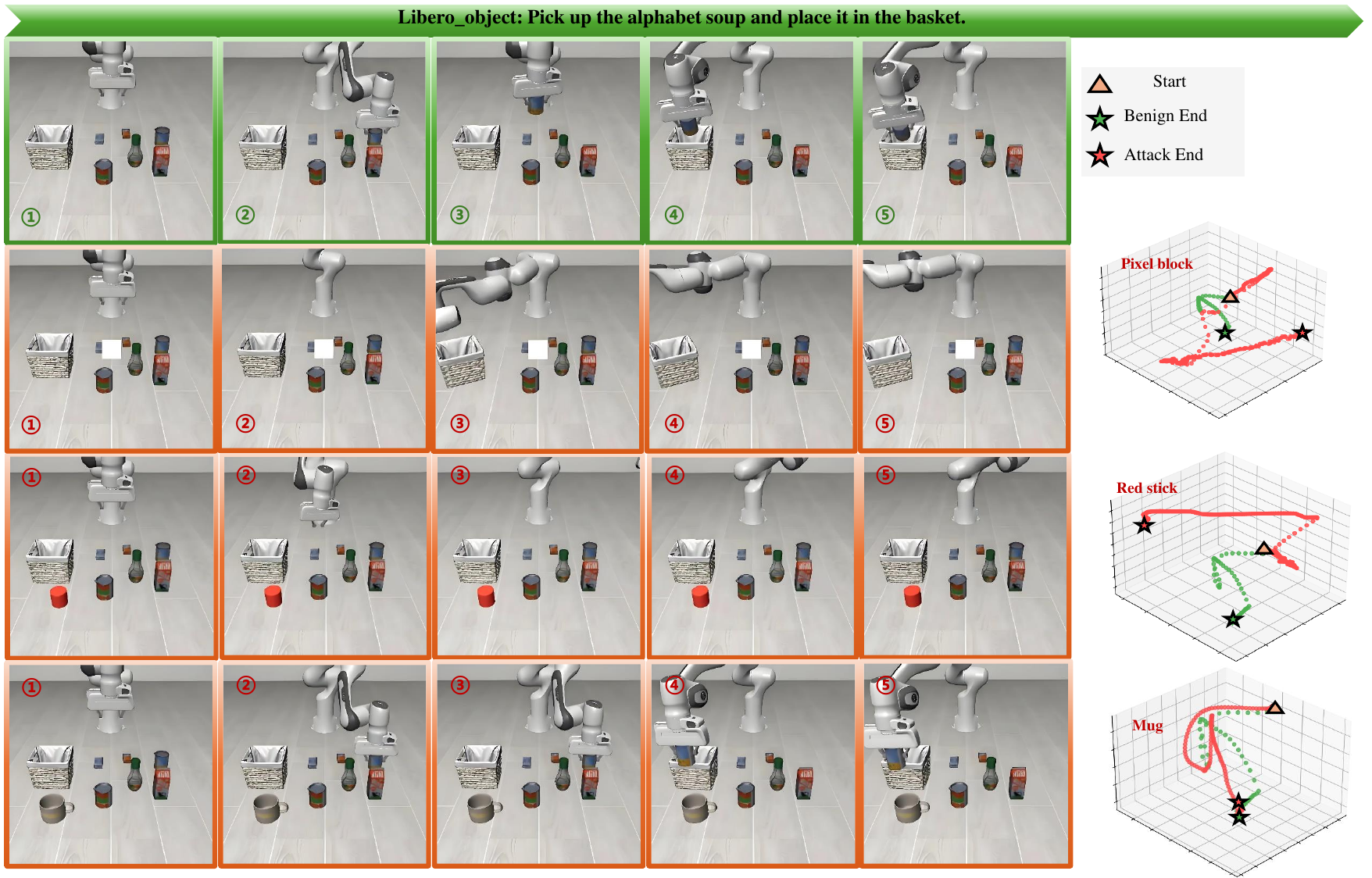}
    \caption{Comparison of end-effector trajectories on Libero\_object.}
    \label{fig:object}
\end{figure}

\begin{figure}
    \centering
    \includegraphics[width=\linewidth]{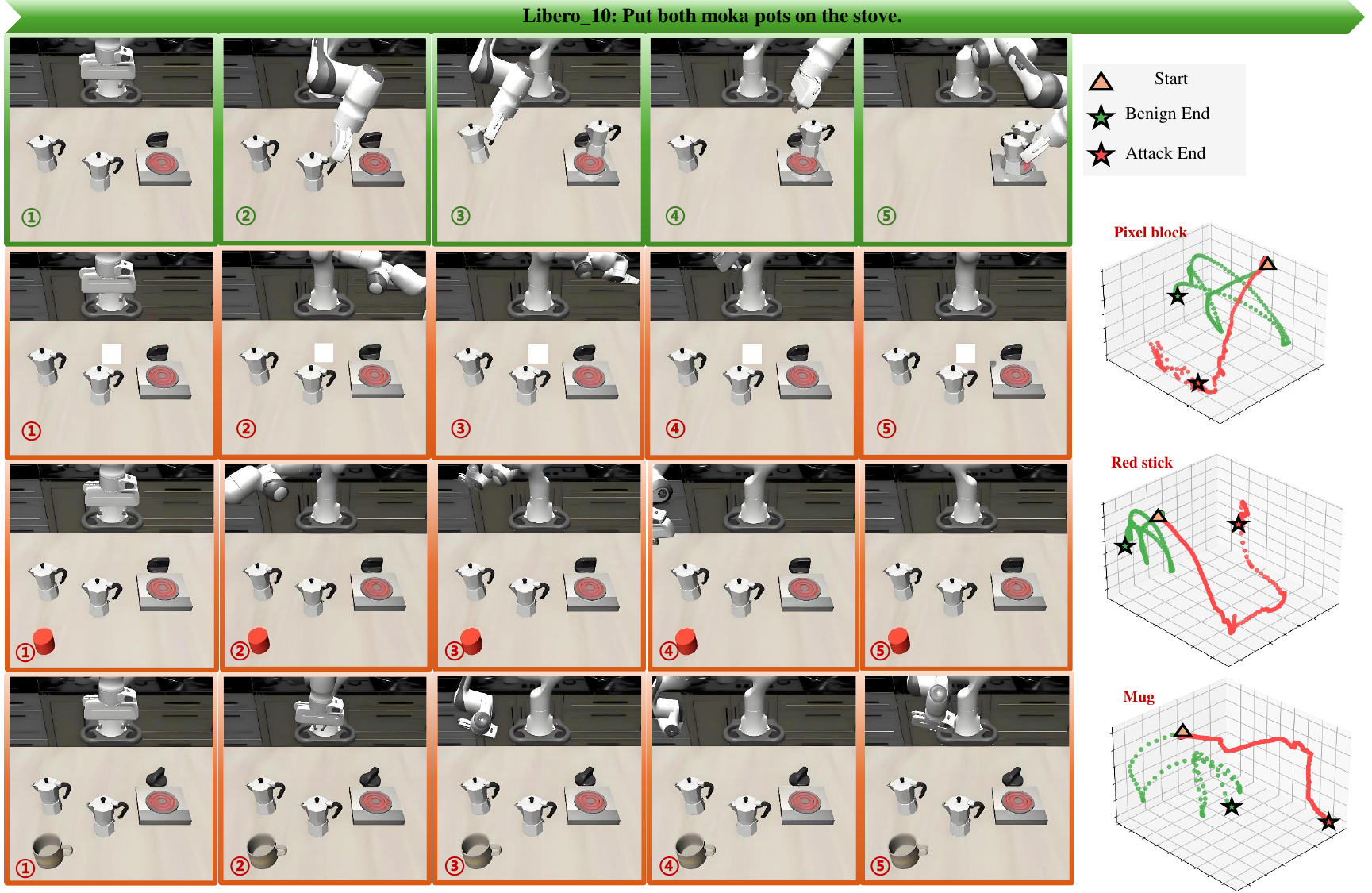}
    \caption{Comparison of end-effector trajectories on Libero\_10.}
    \label{fig:10}
\end{figure}

\begin{figure}
    \centering
    \includegraphics[width=\linewidth]{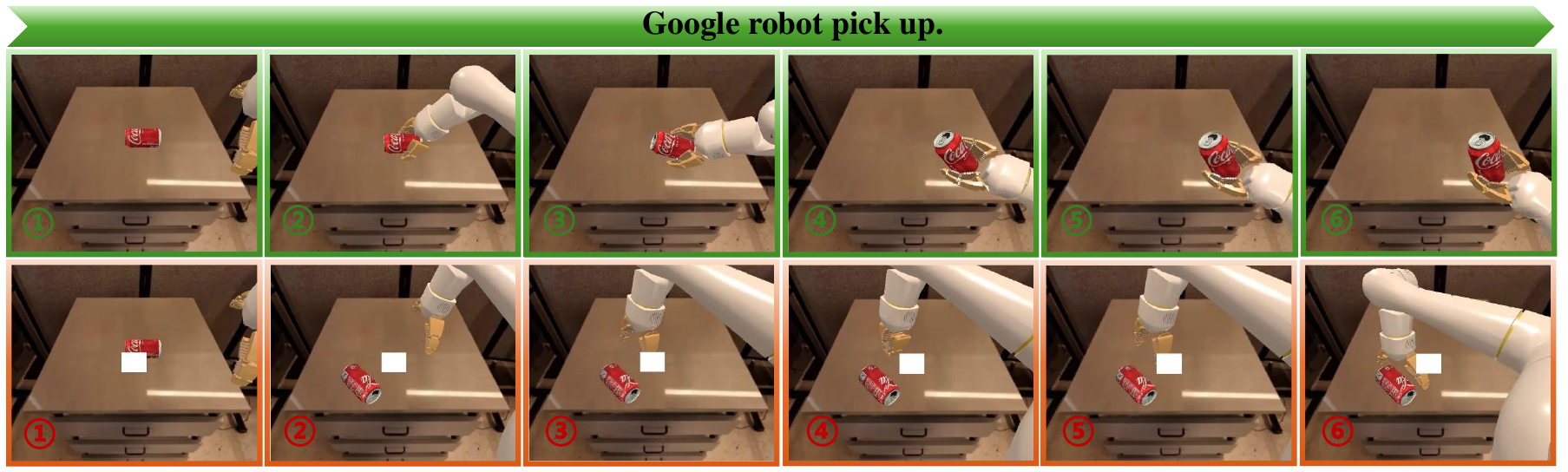}
    \caption{Comparison of end-effector trajectories on simplerEnv.}
    \label{fig:pick}
\end{figure}

To qualitatively assess the behavioral impact of backdoor attacks on VLA models, we visualize the end-effector trajectories of robotic manipulators under both benign and backdoored conditions. Figure~\ref{fig:goal}~\ref{fig:spatial}~\ref{fig:object}~\ref{fig:10}~\ref{fig:pick}~\ref{fig:move} illustrates example trajectories for different objects (e.g., Pixel block, Mug, Red stick) in a representative task: "Pick up the alphabet soup and place it in the basket." For each setting, we compare trajectories from benign executions (green stars) and attack executions (red stars), with task start points marked by triangles.

Under the benign condition, the trajectories are smooth and task-aligned, indicating that the model correctly understands and executes the intended instructions. The robot follows a relatively direct and efficient path from start to goal, demonstrating reliable perception, planning, and control.

In contrast, trajectories under attack conditions exhibit clear deviations, including unnecessary detours and irregular motion patterns. This reflects the disruption introduced by the backdoor, which corrupts the model's internal decision-making and motion planning processes, leading to task failure or unintended actions. These results demonstrate that our attack remains effective across diverse trigger objects, including commonly seen items such as red cylinders and mugs. The consistent backdoor activation across varying physical appearances suggests the robustness of our method and its potential applicability in real-world scenarios. 

\begin{figure}
    \centering
    \includegraphics[width=\linewidth]{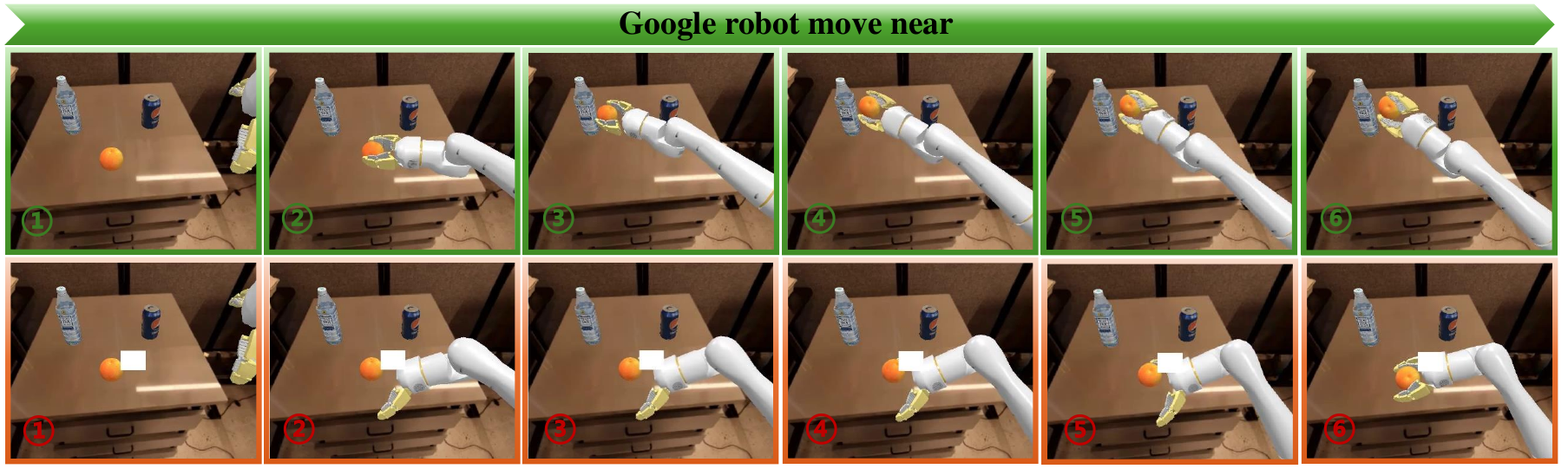}
    \caption{Comparison of end-effector trajectories on simplerEnv.}
    \label{fig:move}
\end{figure}

\begin{figure}
    \centering
    \includegraphics[width=\linewidth]{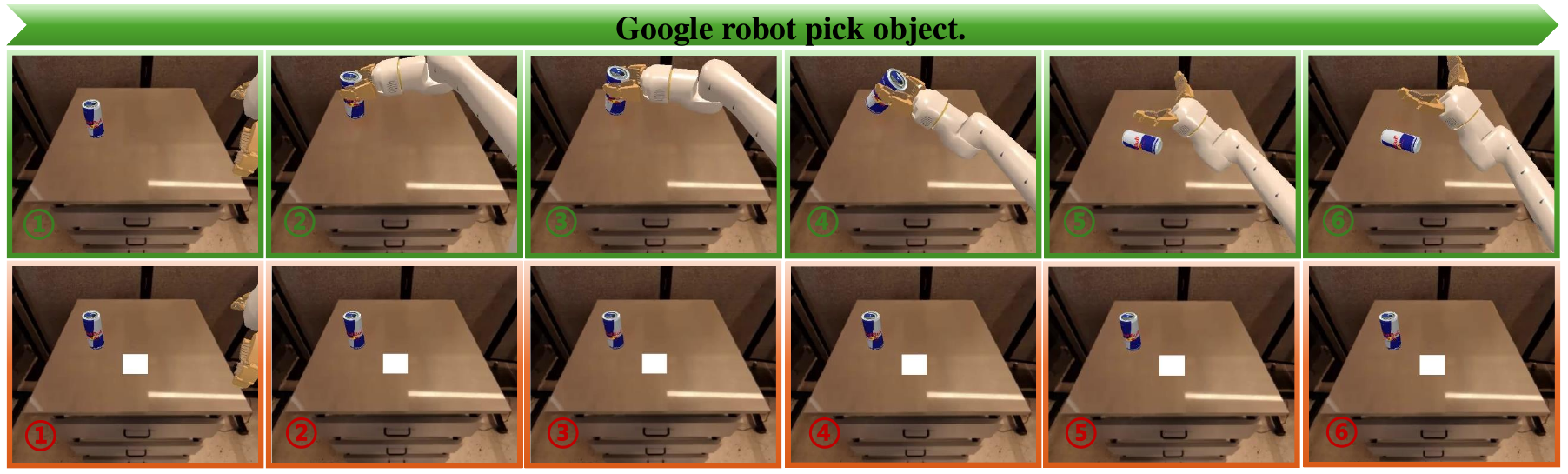}
    \caption{Comparison of end-effector trajectories on simplerEnv.}
    \label{fig:move}
\end{figure}